# Artificial Intelligence-based algorithms in medical image scan segmentation and intelligent visual-content generation - a concise overview

Zofia Rudnicka, Janusz Szczepanski, Agnieszka Pregowska*

Institute of Fundamental Technological Research, Polish Academy of Sciences, Pawinskiego 5B, 02-106 Warsaw, Poland
\*          Correspondence: aprego@ippt.pan.pl; Tel.: +48-22-826-1281 (ext. 412)

**Abstract:** Recently, Artificial Intelligence (AI)-based algorithms have revolutionized the medical image segmentation processes. Thus, the precise segmentation of organs and their lesions may contribute to an efficient diagnostics process and a more effective selection of targeted therapies as well as increasing the effectiveness of the training process. In this context, AI may contribute to the automatization of the image scan segmentation process and increase the quality of the resulting 3D objects, which may lead to the generation of more realistic virtual objects. In this paper, we focus on the AI-based solutions applied in the medical image scan segmentation, and intelligent visual-content generation, i.e. computer-generated three-dimensional (3D) images in the context of Extended Reality (XR). We consider different types of neural networks used with a special emphasis on the learning rules applied, taking into account algorithm accuracy and performance, as well as open data availability. This paper attempts to summarize the current development of AI-based segmentation methods in medical imaging and intelligent visual content generation that are applied in XR. It concludes also with possible developments and open challenges in AI application in Extended Reality-based solutions. Finally, the future lines of research and development directions of Artificial Intelligence applications both in medical image segmentation and Extended Reality-based medical solutions are discussed.

## 1. Introduction

The human brain, a paramount example of evolutionary biological sophistication, transcends its anatomical categorization. Constituted by an estimated 86 billion neurons linked through an intricate web of synapses (ranging in the trillions), it is the epicenter of our cognitive, emotional, and consciousness-related functions [1]. This masterful structure of the central nervous system represents a nexus of myriad neurobiological processes, intricately overseeing sensory input conversion, motoric responses, and advanced cognitive functionalities. As a product of relentless evolutionary adaptations spanning millions of years, the brain epitomizes the apex of neurobiological optimization, synergizing complex neural circuitry with higher-order cognitive undertakings such as cognitive reasoning, emotional homeostasis, and the intricate processes of memory encoding, storage, and retrieval [1-3]. Thus, the human brain is a super-complex system whose functioning and intelligence depend rather on the type of neurons (depending on their role in the brain), their connections, and the way of supplying energy to neurons than the number of neurons [2]. It is an ideal reference model for the foundations of Artificial Intelligence (AI) [3,4].

Thus, processing and analysis of biomedical data for diagnostic purposes is a multidisciplinary field that combines AI, Machine Learning (ML), biostatistics, time series analysis as well as statistical physics and algebra (e.g. graph theory) [3]. Variables derived from biomedical phenomena can be described in several ways and in different domains (time, frequency, spectral values, spaces of states describing the biological system), depending on the characteristics and type of signal. Effective diagnosis of the early stages of the disease, as well as the determination of disease development trends, is a very difficult issue that requires taking into account many factors and parameters. Therefore, the state spaces of biomedical signals are huge and impossible to fully search, analyze, and classify even with the use of powerful computational resources. Therefore, it is necessary to use Artificial Intelligence, in particular, bio-inspired AI methods to limit research to a smaller but significant part of the state space.

Recently, computer-generated three-dimensional (3D) images have become increasingly important in medical diagnostics [5,6]. In particular, Extended Reality (XR) so-called Metaverse is increasingly used in health care and medical education, while it enables the deeper experience of the virtual world, especially through the development of depth perception, including the rendering of several modalities like vision, touch, and hearing [7]. In fact, medical images have different modalities and their accurate classification at the pixel level enables the accurate identification of disorders and abnormalities [8,7]. However, creating a 3D model of organs and/or their abnormalities is time-consuming and is often done manually or semi-automatically[10]. AI can automate this process and also contribute to increasing the quality of the resulting 3D objects [11,12] as well as visual content in the Metaverse [4,13]. To give the users a real sense of visual immersion, the developers should implement virtual objects of high quality [14]. In the context of medicine, it is combined with good quality medical data and their classification/segmentation algorithms with high accuracy, to faithfully reproduce the content in virtual three dimensions.

In this study, we aim to determine existing research gaps in the area of broadly understood medicine, including clinical trials in the application of explainable Artificial Intelligence. For that reason, this paper focuses on the overview of Artificial Intelligence-based algorithms in medical image scan segmentation and intelligent visual content generation in Extended Reality, including different types of neural networks used and learning rules, taking into account mathematical/theoretical foundations, algorithm accuracy, and performance, as well as open data availability. Specifically, we aim to answer the following research questions: can AI-based algorithms be used for the accurate segmentation of medical data?, and how can AI-based algorithms be beneficial in Extended Reality-based technologies?

## 2. Materials and Methods

The methodology of the systematic review was based on the PRISMA Statement which was published in several journals [15] and its extensions: PRISMA-S [16]. We considered recent publications, reports, protocols, and review papers from Scopus and Web of Science databases. The keywords: Artificial Intelligence, Machine Learning, Extended Reality, Mixed Reality (MR), Virtual Reality (VR), Metaverse, learning algorithms, learning rules, signal classification, signal segmentation, medical image scan segmentation, segmentation algorithms, classification algorithms, and their variations. The selected sources were analyzed in terms of compliance with the analyzed topic, and then their contribution to medical image scan segmentation. First, the obtained title and abstract were independently evaluated by the authors. The duplicated records have been removed. Moreover, we have considered the inclusion of criteria-like publication in the form of journal papers, books, and proceedings as well as technical reports. The search was limited to full-text articles in English, including electronic publications before printing. Also, the exclusion criteria like Ph. D. thesis and materials not related to medical image scan segmentation and Artificial Intelligence-based algorithms have been adopted. Subsequently, articles meeting the criteria were retrieved and analyzed. The documents used in this study were selected based on the procedure presented in **Figure 1**. Finally, 162 documents were taken into account. The main limitation of the presented study was the fact that the field of medical image segmentation was lack of the theoretical principles in the considered paper, and lacked critical information concerning development algorithms based on Artificial Intelligence, such as type of the neural network, neuron model, information concerning details in data sets, input and output parameter, learning rules (treating AI-based systems) like a black box.

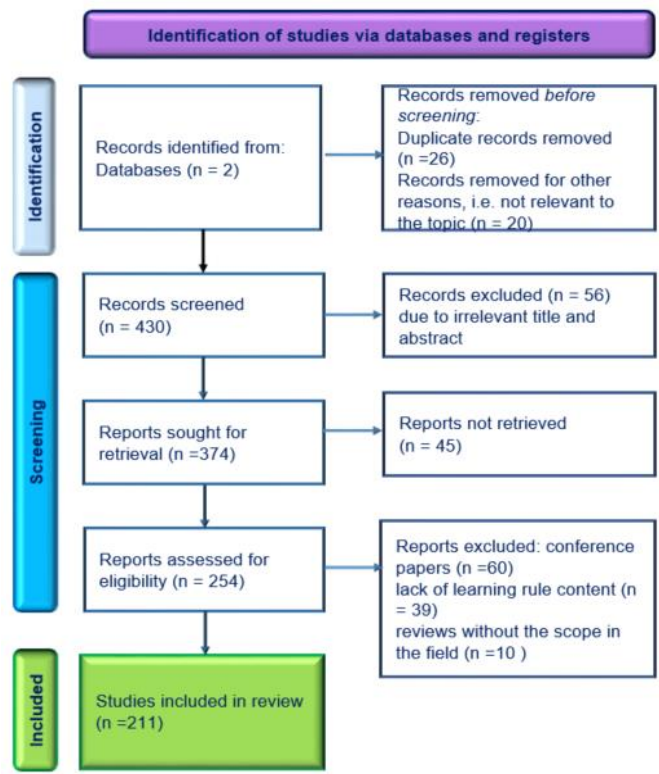

**Figure 1**. The scheme of the methodology of literature review.

In this study, we concentrate on the theoretical foundation of neural communication, the model of neurons, the type of neural networks, and learning rules with a special emphasis on their application in medical image scan segmentation and intelligent visual-content generation. We analyzed the Artificial Neural Networks (ANNs), Convolutional Neural Networks (CNNs), Recurrent Neural Networks (RNNs), Spiking Neural Networks (SNNs) as well as Generative Adversarial Networks (GAN), Graph Neural Networks (GNNs) and Transformers. The first one is made by the simplest neuron model (i.e. perceptions) and can process the information only in one direction. The second one consists of multilayer perceptrons and contains one or more convolutional layers that are responsible for the creation of feature maps, which are subjected to nonlinear processing. RNNs save the output to the processing nodes and feed the result back into the network (bidirectional information processing). The last type is closest to the real nervous system. SNNs transmit the information in the case the membrane potential of a neuron reaches the threshold not in every cycle of propagation like other listed neural networks. Another field that we analyzed in the context of medicine is learning rules, including backpropagation (i.e. in which the weight of the network is calculated according to the chain rule of the partial derivatives of the error function), ANN-SNN Conversion (i.e. transforming SNN networks into ANNs and application of the learning rules that are efficient in ANNs), Supervised Hebbian Learning (i.e. the postulate based on the rule that when the human brain in learning the neurons activates), Reinforcement Learning with Supervised Models (i.e. it enables the monitoring the reaction on the learning rule),

Chronotron (i.e. learning rules which take into account both, spiking neuron and the time of spiking), and biologically inspired network learning algorithms.

## 3. Neural communication

Neurons, which are basic brain building blocks, function as the core computational units of the brain, underpinning the vast expanse of conscious and subconscious processes, and defining our neural identity with each electrochemical interaction [11]. Neurons communicate with other neurons, and non-neuronal cells like muscles and glands by biological connections called "chemical synapses", which are the communication points, at which sending nerve cells called presynaptic neurons, transmit the message to receiving nerve cells called the postsynaptic neuron. The presynaptic neuron releases neurotransmitters, a diverse group of chemicals, into the synaptic cleft (i.e. small gap at which neurons communicate). Following the release, these compounds traverse the synaptic gap, interacting with receptors on the postsynaptic membrane, eliciting a series of intracellular events, potentially leading to the generation of an action potential, a transient depolarizing event propagated along the neuronal membrane.

Since the famous experiments of Adrian [17, 177], it is assumed that in the nervous systems (including the brain), information is transmitted through weak electric currents (on the order of 100 (mV)), in particular employing action potentials (spikes) that are a transient, sudden (1-2 millisecond) change in the membrane potential of the cell/neuron associated with the transmission of information [18]. The stimulus for the creation of an action potential is a change in the electric potential in the cell's external environment. A wandering action potential is called a nerve impulse. In literature [19,20] it is assumed that the sequences of such action potentials, called spike-trains, play a key role in the transmission of information, and the times of appearance of these action potentials play a significant role. Mathematically, such time sequences can be and are modeled in particular after digitalization as trajectories (or their various variants) of certain stochastic processes (Bernoulli, Markov, Poisson, ...) [19,21-27].

## 4. Taxonomy of neural network applied in the medical image segmentation process

The Artificial Neural Networks (ANNs) are constructed with the perceptron neuron model [28] that is based on the binary decision rule. If the linearly weights $w_i$ the sum of the input signals (input vector $x_i$) exceeds the threshold $t_{hr}$ neuron fires (i.e. the output is equal to 1) or if not output is equal to 0.

The basic input function is described as follows

$$f(x) = \begin{cases} 1, & \text{if } w_1x_1 + w_2x_2 + \cdots + w_nx_n \geq t_{hr} \\ 0, & \text{otherwise} \end{cases} \quad (1)$$

The output vector of all neurons in $l$-th layer can be expressed as well as the combination of the linear transformation and non-linear mapping (i.e. ANN activation values) [29].

$$a^l = h(\boldsymbol{W}^l a^{l-1}), i = 1, \dots, M \quad (2)$$

where $W^l$ is the weight matrix between layer $l$ and $l-1$, and $h(\cdot)$ denotes the activation function, in this case, Rectified Linear Unit (ReLU) $f(x) = x^+ = \max(0, x)$ and the vector $a^l$ denotes the output of all neurons in $l$-th layer. The formula (2) has been quoted following the designations in the publication [29]. Neuron models from the Integrate-and-Fire family are among the simplest, however, also the most frequently used. They are classified as spiking models. From a biophysical point of view, action potentials are the result of currents flowing through ion channels in the membrane of nerve cells. The Integrate-and-Fire neuron model [30, 31] focuses on the dynamics of these currents and the resulting changes in membrane potential. Therefore, despite numerous simplifications, these models can capture the essence of neuronal behavior in terms of dynamic systems.

The concept of Integrate-and-Fire neurons is the following: the input ion stream depolarizes the neuron's cell membrane, increasing its electrical potential. An increase in potential above a certain threshold value $U_{thr}$ produces an action potential (i.e. an impulse in the form of Dirac's delta) and then the membrane potential is reset to the resting level. The leaky Integrate-and-Fire (LIF) neuron model [30, 31] is an extended model of the Integrate-and-Fire neuron, in which the issue of time-independent memory is solved by equipping the cell membrane with a so-called leak. This mechanism causes ions to diffuse in the direction of lowering the potential to the resting level or another level $U_0 \to U_{leak} < U_{thr}$. Thus, the third generation of neural networks, i.e. the Spiking Neural Networks (SNN) [32] are mostly based on the LIF, where the membrane potential $U(t)$ is determined by the equation

$$\tau_m \frac{dU}{dt} = -[U(t) - U_{rest}] + R_m I(t) \quad (3)$$

where $\tau_m$ is the membrane time constant of the neuron, $R_m$ is total membrane resistance, and $I(t)$ is the electric current passing through the electrode. The spiking events are not explicitly modeled in the LIF model. Instead, when the membrane potential $U(t)$ reaches a certain threshold $U_{th}$ (spiking threshold), it is instantaneously reset to a lower value $U_{rest}$ (reset potential) and the leaky integration process starts a new one with the initial value $U_r$. To mention just a little bit of realism to the dynamics of the LIF model, it is possible to add an absolute refractory period $\Delta_{abs}$ immediately after $U(t)$ hits $U_{th}$. During the absolute refractory period, $U(t)$ might be clamped to $U_r$, and the leaky integration process is re-initiated following a delay of $\Delta_{abs}$ after the spike. More generally, the membrane potential (3) can be presented as

$$U(t) = \sum_{i=1}^{N} \omega_i \sum_{t_i < t} u(t - t_i) \quad (4)$$

where $u(t)$ is a fixed casual temporal kernel that is an operation that allows scale covariance and scale invariance in a causal-temporal and recursive system over time [33] and $\omega_i, i = 1, \dots, N$ denotes the strength of neuron synapses. Following Equation (2), the neuron's output $m^l(t)$ (membrane potential after the neuron firing) can be described as follows [29]

$$m^l(t) = v^l(t-1) + W^l x^{l-1}(t) \quad l = 1, \dots, N \quad (5)$$

where $v^l$ denotes the membrane potential before the neuron fires, $W^l$ is the weight in $l$-th layer ($l$ denoted layer index), and $x^{l-1}(t)$ is the input from the last layer. Thus, to avoid the loss of information the reset-by-subtraction" mechanism was introduced [34]

$$v^l(t) - v^l(t-1) = W^l x^{l-1}(t) - (H(m^l(t) - \boldsymbol{\theta}^l)\theta^l) \quad (6)$$

where $v^l(t)$ is membrane potential after firing, $m^l(t)$ − membrane potential before firing, $H(m^l(t) − \theta^l)$ refers to the output spikes of all neurons, and $\boldsymbol{\theta}^l$ is a vector of the firing threshold $\theta^l$. There are also some applications of the concepts of the meta-neuron model in SNNs [35]. The main differences between the LIF neuron and meta neurons stay in the integration process, where meta neurons use a 2nd-order ordinary differential equation and an additional hidden variable. The basic differences between ANN and SNN (taking into account the type of neuron models) are presented in **Figure 1**.

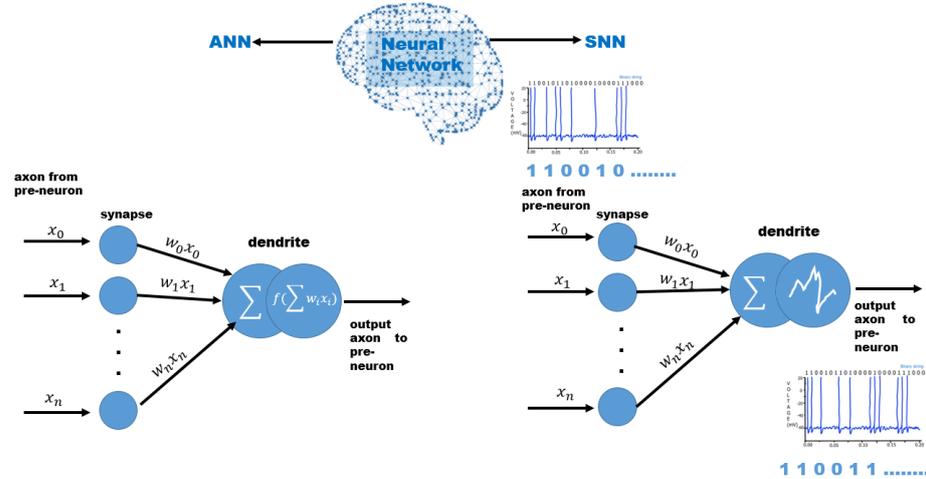

**Figure 2**. The scheme of the basic differences between ANN and SNN takes into account the type of neuron models.

*4.1. Convolutional Neural Network*

The most commonly used deep neural network (DNN) in medical image classification is the two-dimensional (2D) Convolutional Neural Network (CNN) [36,37]. In **Figure 2**. The basic scheme of the SNN is presented. Its principle of operation is based on linear algebra, in particular matrix multiplication. CNNs consist of three types of layers: a convolutional layer, a pooling layer, and a fully connected layer. In fact, most computations are performed in the convolutional layer or layers. The image (pixels) is converted into binary values and patterns are searched. Every convolutional layer operates a dot product between two matrices, namely one matrix is a set of learnable parameters (kernel), and the second matrix is a limited part of the receptive field. Each subsequent layer contains a filter/kernel that allows you to classify features with greater efficiency. A pooling layer reduces the number of parameters in the input, which causes the loss of part of the information calculated in the common layer/layers, however, it allows for improvement in the efficiency of the CNN network. This operation is performed by sliding windows [38]. Next, the output of these two layers is transformed into a one-dimensional vector, i.e. input to the fully connected layer. In this last type of layer, image classification based on the features extracted in the previous layers is performed, i.e. the object in the image is recognized. The output $y_{i,j}^{(k)}$ from CNN can be described as follows

$$y_{i,j}^{(k)} = \sigma(\sum_{l=1}^{L} \sum_{m=1}^{M} x_{i+l-1,j+m-1}^{(l)} w_{l,m}^{(k)} + b^{(k)}) \quad (7)$$

where $x_{i,j}^{(l)}$ denotes input to the network at the spatial location $(i,j)$, $\sigma$ is the activation function, $w_{l,m}^{(k)}$ is the weight of the $m$th kernel at the $l$th channel producing the $k$th feature map, and $b^{(k)})$ is the bias for $k$th feature map.

In the case of large datasets, CNN achieves high efficiency and is resistant to noise [39]. The crucial disadvantages of CNNs in image processing are high computational requirements and difficulties in achieving high efficiency in the case of small datasets (i.e. if the dataset is too small the network may overfit to training data, and poorly recognize new data).

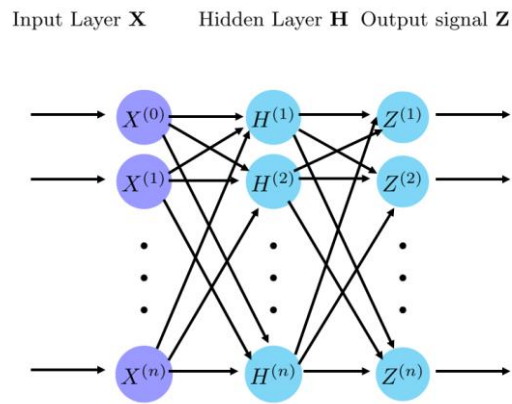

**Figure 3**. The basic scheme of the simple Convolutional Neural Network.

*4.2. Recurrent Neural Network*

Another neural network commonly applied in medical data analysis is the Recurrent Neural Network (RNN) [40]. In the **Figure 3**. The basic scheme of the RNN is presented. This type of network contains at least one feedback connection. The output of RNN can be expressed as [41]

$$y_i = W_{hy}\,\mathcal{H}(W_{hh}h_{i-1} + W_{xh}x_i + b_h)h_i + b_y \qquad (8)$$

where $x_i$, $i = 1, ..., T$ is the input sequence of T states $(x_i, ..., x_T)$ with $x_i \in \mathbb{R}^d$, $W_{xh}$, $W_{hy}$, $W_{hh}$ denotes weight matrices, $b_h$, $b_y$ are bias vectors, and $\mathcal{H}$ is the non-linear activation function, for example, ReLU, Sigmoid $f(x) = \frac{1}{1+e^{-x}}$, Tanh Function (Hyperbolic Tangent) $f(x) = \frac{e^x - e^{-x}}{e^x + e^{-x}}$. The network operation is recursive since the hidden layer state depends on the current input and the previous state of the network. Thus, the hidden state $h_{i-1}$ is the memory of past inputs.

Thus, the RNN can operate on the sequential dataset and has an internal memory. It may have many inputs. However, RNNs exhibit learning-related problems, namely vanishing gradients (i.e. in the case of small gradients the updates of parameters are irrelevant) or exploding gradients (i.e. superposition of large error gradients leading to large parameter updates). These contribute to the long training process, low level of accuracy, and low network performance.

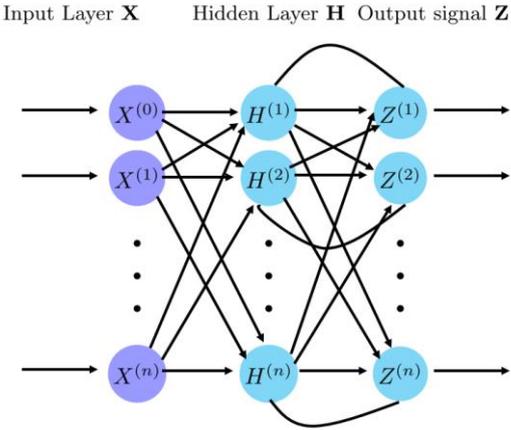

**Figure 3**. The basic scheme of the simple Recurrent Neural Network.

*4.3. Spiking Neural Networks*

Besides the Artificial Neural Networks, i.e. CNNs, and RNNs, one can also be applied to the medical signals bio-inspired neural networks like Spiking Neural Networks [41,42]. In the **Figure 4**. The basic scheme of the SNN is presented. SNNs encode information taking into account spike signals, and shells are promising in effectuating more complicated tasks, while the more spatiotemporal information is encoded with spike patterns [43]. They are mostly based on the LIF neuron model. SNNs were formulated to map organic neurons, i.e. the appearance of the presynaptic spike at synapse triggers the input signal $i(t)$ (the value of the current) that in the simplified cases can be written as follows

$$i(t) = \int_0^\infty S_j(s-t)\exp(\frac{-s}{\tau_s})\mathrm{d}s \qquad (9)$$

where $\tau_s$ denotes synaptic time constant, $S_j$ is a presynaptic spike train, $t$ is time [44]. In contrast, the majority of DNNs do not take into account temporal dynamics [45]. In fact, SNNs show promising capability in playing a similar performance as living brains. Moreover, the binary activation in SNNs enables the development of dedicated hardware for neuromorphic computing [46]. The potential benefits are low energy usage and greater parallelizability due to the local interactions.

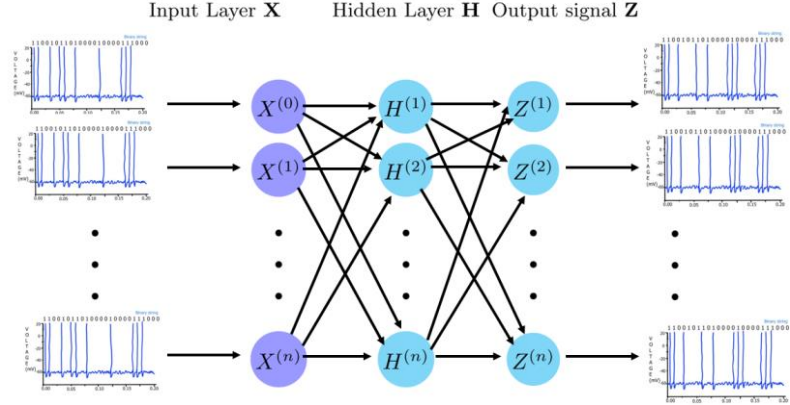

**Figure 4**. The basic scheme of the simple Spiking Neural Network.

**5. Learning algorithms**

The heart of Artificial Intelligence is its learning algorithms. At their core, strive to automate the learning process, enabling machines to recognize patterns, make decisions, and predict outcomes based on data. Their design is often a balance between theoretical rigor and practical applicability. While mathematics and statistics provide the foundation, translating these into algorithms that can operate on vast and diverse datasets requires creative programming skills [22]. One can distinguish many types of network training algorithms [47]. Below we briefly discuss the most important of them, taking into account the theoretical foundations.

5.1 Back Propagation Algorithm

The most commonly used learning algorithm is the back propagation (BP) algorithm. Ititers overweight optimizations via error propagation in the neural networks. BP plays a pivotal role in enabling neural networks to recognize complex and non-linear patterns from large datasets [23,48, 49]. From the mathematical point of view, it is a calculation of the cost function, which minimizes the calculated error of the output using gradient descent or delta rule [50]. It can be split into three stages: forward calculation, backward calculation, and computing the updated biases and weights. The input to the hidden layer $H_j$ is the weighted sum of the outputs of the input neurons and can be described as [51]

$$H_j = b_{in} + \sum_{i=1}^{n} x_i w_{ij} \quad (10)$$

where $x_i$ is the input to the network (input layer), $n$ is the number of neurons in the input layer, $b_{in}$ is the bias input layer, and $w_{ij}$ denotes the weight associated with the $i$-th input neuron and the $j$-th hidden neuron. The output $y_k$ is as follows

$$y_k = b_h + \sum_{j=1}^{m} w_{jk} F(H(j)) \quad (11)$$

where $F(H(j))$ is a transfer function, $k$ is the number of neurons in the hidden layer, and $b_h$ is the bias of the hidden layer. The most commonly used transfer function is the sigmoid transfer function $F(H(j)) = \frac{1}{1+e^{-(H(j))}}$. The back propagation algorithm is especially effective when used in multi-layered neural architectures such as feed-forward neural networks, convolutional neural networks, and recurrent neural networks [26]. In image recognition, CNNs, energized by BP, can independently identify hierarchical features, from basic edges to detailed structures. Similarly, RNNs, amplified by BP, are adept at sequence-driven tasks like machine translation or speech recognition, as they incorporate previous data to influence present outputs. It is one of the most effective deep learning methods. However, BP requires large amounts of data and enormous computational efforts.

5.2 ANN-SNN Conversion

Artificial Neural Networks and Spiking Neural Networks are both computational models inspired by biological neural networks. While ANNs have been the mainstream for most deep learning applications due to their simplicity and effectiveness, SNNs are gaining traction because they mimic the behavior of real neurons more closely by using spikes or binary events for communication. To obtain a similar accuracy of the SNN-based algorithm as the algorithm using ANN, for example, the BP-type training rule consumes a lot of hardware resources. The already existing platforms have limited optimization possibilities. Thus, the conversion of ANNs to SNNs seeks to harness the energy efficiency and bio-realism of SNNs without reinventing the training methodologies [28], while it is based on the ReLU activation function and LIF neuron model [52]. The basic principle of the conversion of ANNs to SNNs is mapping the activation value of the ANN neuron to the average postsynaptic potential (in fact, firing Rate) of SNN neurons, and the change of the membrane potential (i.e. the basic function of spiking neurons) can be expressed by the combination of the Equation (2) and Equation (6)[29]

$$v^l(t) - v^l(t-1) = W^l x^{l-1}(t) - s^l(t)\theta^l \quad (12)$$

Here $s^l(t)$ refers to the output spikes of all neurons in layer $l$ at time $t$.

Tuning the right thresholds is paramount for the SNN to effectively and accurately represent information. Incorrectly set thresholds could lead to either too frequent or too rare spiking, potentially affecting the accuracy of the SNN post-conversion [35]. On the other hand, the neuromorphic hardware platforms that support SNNs natively can primarily offer energy efficiency benefits by converting ANNs to SNNs. Due to their event-driven nature, SNNs can be more computationally efficient [36]. However, the challenge lies in maintaining accuracy post-conversion. Some information might be lost during the transition, and not all ANN architectures and layers neatly convert to their SNN equivalents. The conversion from ANNs to SNNs is a promising direction, merging the advanced training methodologies of ANNs with the energy efficiency of SNNs. As we delve deeper into the realm of neuromorphic computing, this conversion process will play a pivotal role in bridging traditional deep learning with biologically-inspired neural models [37, 38].

5.3 Supervised Hebbian Learning (SHL)

Taking into account Artificial Intelligence, Supervised Hebbian Learning (SHL)can be described as a general methodology for weight changes [53]. Thus, this weight increases when two neurons fire at the same time, while it decreases when two neurons fire independently. According to this rule, the change in weight can be written

$$\Delta w = \eta(t^{out} - t^d) \quad (13)$$

where $\eta$ is the learning rate (in fact, the small scalar that may vary with time, $\eta > 0$), $t^{out}$ the actual time of the postsynaptic spike, while $t^d$ is the time of firing of the second presynaptic spike [54, 55]. The crucial disadvantage of Hebbian learning is the fact that when the number of hidden layers increases the efficiency decreases, while in the case of 4 layers is still competitive [56].

5.4 Reinforcement Learning with Supervised Models

According to the additional constraints in the SHL rule, Reinforcement Learning with Supervised Models (ReSuMe) was proposed [54]. ReSuMe, is a dynamic hybrid learning paradigm. It effectively combines the resilience of Reinforcement Learning

(RL) with the precision of Supervised Learning (SL). This fusion empowers ReSuMe to leverage feedback-driven mechanisms inherent in RL and take advantage of labeled guidance typical for SL [37,38, 39]. The difference between SHL is that the learning signal is expected not to have or have a marginal direct effect on the value of the postsynaptic somatic membrane potential [57], thus the synaptic weights are modified as follows

$$\frac{d}{dt}w_{ji}(t) = a[S_d(t) - S_j(t)]\bar{S}_i(t) \quad (14)$$

where $a$ denoted learning rate, $S_d$ is desired/targeted spike train, $S_j(t)$ is the output of the network (spike train), and $\bar{S}_i(t)$ expresses the low-pass filtered input spike train. ReSuMe guided one of its most salient features exploration. By leveraging labeled data *via* SL, ReSuMe can effectively steer RL exploration, ensuring agents avoid falling into the trap of suboptimal policies. The hybrid nature of ReSuMe also grants it a unique resilience, especially in the face of noisy data or in reward-scarce environments. Moreover, its adaptability is noteworthy, making it an ideal choice for tasks that combine immediate feedback (through SL) with long-term strategic maneuvers (through RL). However, like all things, ReSuMe is not without challenges. A potential bottleneck in ReSuMe is computational complexity, as managing both RL and SL can sometimes strain computational resources. Another challenge is the precise tuning of the $a$ coefficient. The key is to find a balance where neither RL nor SL overly dominates the learning process. By melding immediate feedback from supervised learning with a deep reinforcement learning strategy, ReSuMe establishes itself as a formidable tool in Machine Learning [49, 50, 52].

5.5 Chronotron

The Chronotron, by its essence, challenges and reshapes our understanding of how information can be encoded and processed in neural structures [50, 55]. Traditional neural models have predominantly focused on the spatial domain, emphasizing the architecture and interconnections between neurons. While this spatial component is undeniably critical, it offers only a part of the full informational symphony that the brain plays. Just as the rhythm and cadence of a song contribute as much to its essence as its melody, in the vast theater of the brain, timing is not just a factor; it is a storyteller in its own right. The brilliance of the Chronotron lies in its ability to discern and respond to this temporal narrative. Unlike its counterparts, which often treat time as a secondary parameter, the Chronotron places it center stage. As a consequence, it acknowledges and leverages the intricate interplay of spatial and temporal dynamics in neural computation. This means that it doesn't just consider which neurons are firing, but also pays meticulous attention to when they fire concerning one another. Thus, the membrane potential is

$$u(t) = \eta(t) + \sum_j w_j \sum_{t_j^f \leq t} \varepsilon_j(t, t_j^f) \quad (15)$$

where the models the $\eta$ model's refractoriness is caused by the past presynaptic spikes, $w_j$ is the synaptic efficacy, $t_j^f$ is the time of appearance of the $f$-th presynaptic spike on the $j$ synapse, $\varepsilon_j(t, t_j^f)$ denotes normalized kernel [58]. When $u(t)$ reaches the threshold level, a spike is fired. And $u(t)$ is reset to the value of reset potential. In this approach, it is crucial to find the appropriate error functions, i.e. such an error function that enables the minimization with a gradient descent method [59]. The advantage of this learning rule is the fact that it uses the same coding for inputs and outputs. Chronotron's hallmark, its granularity, can sometimes surge computational demands, especially during intense training. And like many cutting-edge neural frameworks, harnessing Chronotron's full potential can be intricate, necessitatin' fine-tuned parameters and rich, well-timed data.

5.6 Bio-inspired Learning Algorithms

Brain-inspired Artificial Intelligence approaches, in particular spiking neural networks, are becoming a promising energy-efficient alternative to traditional artificial neural networks [60]. However, the performance gap between SNNs and ANNs has been a significant obstacle to the wild SNNs application (applicable SNNs). To fully use the potential of SNNs, including the detection of the non-regularities in biomedical signals, and designing more specific networks, the mechanisms of their training should be improved, one of the possible directions of development is the bio-inspiring learning algorithms. Below we briefly discuss the most important of them.

5.6.1 Spike Timing Dependent Plasticity

Spike Timing Dependent Plasticity (STDP) is rooted in the idea that the precise timing of neural spikes critically affects changes in synaptic strength [61]. This principle highlights the intricate dance between time and neural activity, showcasing the dynamics of our neural circuits. This biologically plausible learning rule is a timing-dependent specialization of Hebbian learning (13) [62]. STDP shed light on the intricate interplay between timing and synaptic modification. It is based on the change in synaptic weight function

$$\Delta W = \eta(1+\zeta)H(W; t_{pre} - t_{post}) \quad (16)$$

where $\eta$ denotes the learning speed, $\zeta$ is Gaussian white noise with zero mean, while $H(W; t_{pre} - t_{post})$ is the function, that determines the long-term potentiation (LTP, ie. presynaptic and postsynaptic neurons emit a high rate) and depression (LTD, i.e. presynaptic neurons emit a high rate) in the time window $t_{pre} - t_{post}$ [63]

$$H(W; t_{pre} - t_{post}) \begin{cases} a_+(W)\exp(-\frac{|t_{pre}-t_{post}|}{\tau_+}) & \text{for } t_{pre} - t_{post} < 0 \\ -a_-(W)\exp(-\frac{|t_{pre}-t_{post}|}{\tau_-}) & \text{for } t_{pre} - t_{post} > 0 \end{cases} \quad (17)$$

where $a(W)$ is a scaling function that determines the weight dependence, while $\tau$ denotes the time constant for depression [61-63]. STDP's significance is underpinned by its numerous advantages. Chiefly, it offers a biologically authentic model by 'mimicking the temporal dynamics observed in real neural 'systems. Furthermore, its event-centric nature promotes unsupervised learning, enabling networks to autonomously adjust based on the temporal patterns present in input data. This time-based sensitivity equips STDP to adeptly process data with spatiotemporal attributes and detect intricate temporal relationships within neuronal signals [64, 65]. However, STDP is not without its complexities. A prominent challenge is the fine-tuning of parameters. The exact values assigned to constants like $a(w)$ and $\tau$ can substantially dictate the behavior and efficacy of STDP-informed networks. Balancing these values requires a meticulous approach. Moreover, the precision demanded by STDP's time-centric nature often calls for higher computational rigor, especially within simulation contexts. STDP stands as a testament to the elegance and intricacy of neural systems. By emphasizing the role of spike timing, STDP offers a vivid depiction of how synaptic interactions evolve [66, 67].

### 5.6.2 Spike-Driven Synaptic Plasticity

Spike-Driven Synaptic Plasticity (SDSP) offers the ability to elucidate the causality in neural communication. It operates on a fundamental principle: the sequence and timing of spikes determine whether a synapse strengthens or weakens. If a neuron consistently fires just before its downstream counterpart, it's a strong indication of its influential role in the latter's activity. This "pre-before-post" firing often leads to synaptic strengthening, cementing the relationship between the two neurons. Conversely, if the sequence is reversed, with the downstream neuron firing before its predecessor, the connection may weaken, reflecting a lack of causal influence [68, 69]. This causative aspect of SDSP provides valuable insights into the learning mechanisms of the brain. It suggests that our neural circuits are continually evolving, adjusting their connections based on the flow of spike-based information. Such adaptability ensures that our brains remain receptive to new information, enabling us to learn and adjust to ever-changing environments. Moreover, SDSP emphasizes the significance of precise spike timing. In the realm of neural computation, milliseconds matter. Small shifts in spike timing can change a synapse's fate, showcasing the brain's precision and sensitivity. This meticulousness in spike-driven modifications underscores the importance of timing in neural computations, hinting at the brain's capacity to encode and process temporal patterns with remarkable accuracy [70-73]. In this learning rule the changes in synaptic weights can be expressed as [64]

$$\Delta w = \begin{cases} \eta^+ + e^{-|\Delta t|/\tau^+} & \text{if } \Delta t > 0 \\ \eta^- + e^{-|\Delta t|/\tau^-}, & \text{otherwise} \end{cases} \quad (18)$$

where $\eta_+ > 0$ and $\eta_- < 0$ denotes the learning parameters, $\tau_+$ and $\tau_-$ are time constraints, and $\Delta t$ is the difference between post- and pre-synaptic spikes. This representation, while streamlined, encapsulates the principle that the mere presence of a spike can induce modifications in the synaptic weight, either strengthening or weakening the connection based on the specific neural context and the directionality of the spike's influence [70].

The appeal of Spkie-Driven Synaptic Plasticity is manifold its primary virtue is its biological relevance. Focusing on individual spike occurrences mirrors the granular events that take place in real neural systems. Such an approach facilitates the modeling of neural networks in scenarios where individual spike occurrences are of paramount importance. Furthermore, by anchoring plasticity on singular events, this model is inherently suitable for real-time learning and rapid adaptability in dynamic environments [Błąd! Nie można odnaleźć źródła odwołania.].

A crucial challenge lies in the accurate capture and interpretation of individual spikes, especially in densely firing neural environments. Moreover, the plasticity model's sensitivity to' single events 'means that it can' be susceptible to noise, requiring sophisticated filtering mechanisms to discern genuine learning events from spurious spikes. SDS elucidates the profound influence of singular neuronal events on the grand tapestry of neural learning and adaptation [75].

### 5.6.3 Tempotron Learning Rule

One of the most interesting biological-inspired learning algorithms is the tempotron principle [65, 76, 77]. It is designed to adapt synaptic weights based on the temporal precise patterns of incoming spikes, rather than only the frequency of such spikes. While traditional neural models might emphasize synaptic weights or connection topologies, tempotron underscores that the 'when' of a neural event can be as informative, if not more so, than the 'where' or 'how often' [78-80]. The tempotron learning rule is based on the LIF neuron model. It fires when the membrane potential described by Equation (4) exceeds the threshold (binary decision). Thus, one can define the potential of the neuron's membrane as a weighted sum of postsynaptic potentials (PSPs) from all appearance spikes [77]

$$v(t) = \sum_i \omega_i \sum_{t_i} K(t - t_i) + V_{rest} \quad (19)$$

where $\omega_i$ denotes synaptic efficacy, $t_i$ is the firing time of the $i$th afferents, $V_{rest}$ is resting potential, and $K$ is the normalized PSP kernel

$$K(t - t_i) = V_0 \left( \exp\left(\frac{-(t - t_i)}{\tau_m}\right) - \exp\left(\frac{-(t - t_i)}{\tau_s}\right) \right) \quad (20)$$

where $\tau_m$ is the decay time constant of membrane integration, while $\tau_s$ denotes the decay time constant of synaptic currents. While the $V_0$ normalized the PSP that the maximum kernel value is equal to 1. The neuron is fired when the value of the potential of the neuron's membrane described by Equation (19) is greater than the value of the firing threshold. Next, the potential of the neuron's membrane described by Equation (19) smoothly decreases to the value of $V_{rest}$. In the case of the segmentation/classification task, the input to the neuron may belong to one of two classes, namely $P^+$ when a stimulus occurs (i.e. pattern is presented) the neuron should fire), and $P^-$ when the pattern is presented neuron should not fired. Each input consists of $N$ spike trains. In turn, the tempotron learning rules are as follows

$$\Delta\omega_i = \lambda \sum_{t_i<t_{max}} K(t_{max} - t_i) \quad (21)$$

where $t_{max}$ is the time when the potential of the neuron's membrane (19) reaches a maximum value. While $\lambda$ is the constant that is greater than zero in the case of $P^+$, and smaller than zero in the case $P^-$. In this operation, tempotron introduces gradient-decent dynamics, i.e. minimizing the cost function for each input pattern measures the maximum voltage that is generated by the erroneous patterns. In comparison to the STDP learning rule, tempotron can make the appropriate decision under a supervisory signal, by tuning fewer parameters than STDP. Thus, tempotron uses LTP and LTD mechanisms like STDP. The advantage of the tempotron learning rule is the speed of learning.

## 6. Neural networks and learning algorithms in the medical image segmentation process

Image segmentation has a crucial role in creating both, medical diagnosing supported by image analysis and virtual object creation like the medical digital twin (DT) of organs [66,67], holograms of the human organs [81, 82], and virtual medical simulators [68, 83]. One can split the image segmentation process into semantic segmentation (i.e. assigning a label or category to each pixel), instance segmentation (i.e. identifying and separating individual objects in an image and assigning a label to it), and panoptic segmentation (i.e. more complex tasks, which involves the two segmentations above) [77, 78]. The application of AI enables to increase in the efficiency and speed of these processes [84]. In Table 1. the comparison of the AI-based algorithms applied in medical image scan segmentation taking into account the neuron model, the type of neural network, learning rule, and biological plausibility is shown. It turned out that the most commonly used in image segmentation are CNNs, in particular, Unet architecture and its variations [71,72. 74, 75, 85]. In [73] the authors modified this neural network structure by adding dense and nested skip connections (UNet++), while [178] added the residual blocks and attention modules to enable the network to learn deeper features and increase the effectiveness of segmentation. To connect the efficiency of segmentation with access to global semantic information, often CNNs are combined with transformer blocks [85-87]. Another CNNs-based algorithm commonly used in medical image segmentation is You Only Look Once (YOLO), which is open-source software used under the GNU General Public License v3.0 license [88, 179]. It uses one fully connected layer, the number (depending on the version) of convolution layers that are pre-trained with the CNN (YOLO v1 ImageNet, YOLO v2 Darknet-19, YOLO v3 Darknet-53, YOLO v4 CSPNet, YOLO v5 EfficientNet, YOLO v6 EfficientNet-L2, YOLO v7 ResNET, YOLO v8 RestNet), and pooling layer. The algorithm divides the input in the form of a photo into specific segmentations and then uses CNN to generate bounding boxes and class predictions. Recently, in image classification, SNN has become more popular [78, 79] due to its low power consumption. However, SNN training rules require refinement to achieve ANN accuracy. However, the development of an efficient, automatic segmentation procedure is of high importance [207].

Recently, Transformer Networks that were designed to machine translation (Natural Language Processing task) have been applied in the field of image processing, including medical image processing [180]. This architecture is based on network normalization feed-forward network and residual structures (namely Multi-Head Attention (MHA) and Positionwise Feed-Forward Networks), while it does not contain any convolutions [181]. Such an architecture enables them to achieve a powerful ability to represent long-term receivables. Thus, the architecture of transformers in the field of computer vision contains only vision transformers (ViT) and Swin transformers [182, 183]. The MHA has multiple attention modules which learn different aspects in different subspaces. In [184] was shown that Transformers may have a higher level of efficiency in the field of image processing compared to the CNNs taking into account learning which is applied to large datasets. To increase the applicability and accuracy of transformers in the area of image processing, data augmentation and regularization strategies are used, among others [185]. On the other hand, vision transforms do not contain inductive biases. Also, the combination of CNNs and Transformers was applied to image processing [186], which contributed to reducing the consumption of computing resources and training time [187, 188]. The main disadvantages of Transformers are the need for commitment of large amounts of computational resources and the requirements of the long training time.

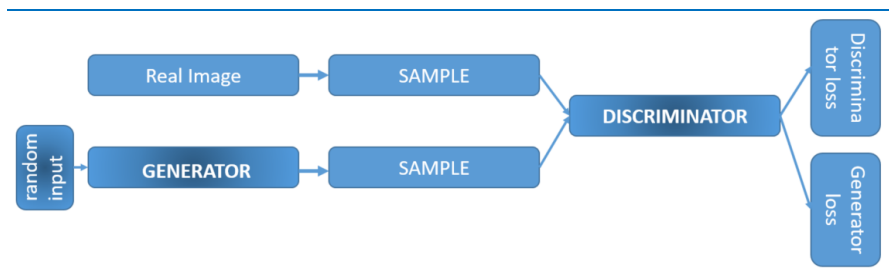

**Figure 5**. The basic scheme of the simple Generative Adversarial Network.

Also, Generative Adversarial Networks are applied to medical image fusion [190]. This type of approach divided the neural networks into two parts: generators (learn to generate reliable data. The generated instances become negative examples for training the second part of the network) and discriminators (i.e. binary classifiers that learn to distinguish generators from real data, see Figure 5. The discriminator presents a penalty for generating meaningful results. The output of the generator is connected directly to the input of the discriminator. In back-propagation, the discrimination classification determines the signal that the generator uses to update the weights. In fact, GANs are the paris of CNNs that are connected adversarial. The difference between them is their approach to getting results. For example, in [191, 192] GAN was successively applied to the segmentation of the blood vessels of the retinal and the coronary with high accuracy. However, centralized training algorithms can potentially mishandle sensitive information such as medical data. Additionally, GANs have significant security issues, such as vulnerabilities that exploit the real-time nature of the learning process to generate prototype samples of private training sets [194]. Also, the use of deep neural networks such as CNN and GAN is limited by the need to have large annotated data sets, which is quite a challenge, especially in medicine [193].

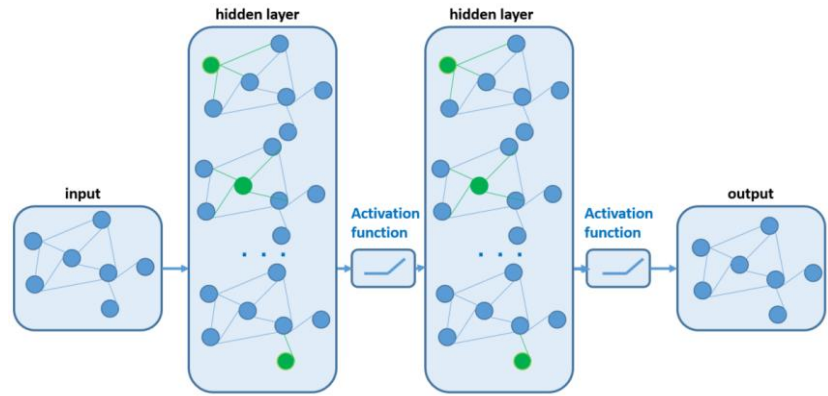

**Figure 6**. The basic scheme of the simple Graph Neural Network.

All the above solutions are based on Euclidean space data that have fixed dimensions. However, data can be also presented in non-Euclidean space (i.e. graphs, namely a set of objects (vertices) and relationships between these objects (edges)) [195]. This kind of data has dynamical dimensions, i.e. the input data do not have to be in particular order like in the case of Euclidean space data [196]. Thus, in the field of medical data processing also occur irregular spatial patterns that may be important for the diagnostics point of view. The analysis of these patterns is a challenge that is proposed to be solved by applying Graph Neural Networks [197, 202, 204]. GNNs are based on the convolution operation on the graphs, see Figure 6. One disadvantage of GNNs is the fact that they are strongly dependent on the geometry of the graph. Consequently, the neural network must be trained every time data is added. In the context of large image diagnostics, this can make a less practical approach. Also, the low computational speed of GNNs taking into account medical data processing contributes to the fact that GNNs need further development for practice applications. As a solution to improved calculation efficiency, a framework for inductive representation learning on large graphs i.e., GraphSAGE was proposed [198]. Thus, GNNs can expand the possibilities of training CNNs on non-grid data [199]. In the field of medical image segmentation, GNNs find especially application in tissue semantic segmentation in histopathology images [203, 205, 206]. In the case of tumor segmentation, the application of CNN leads to the number of parameters that contribute to the high computational complexity. Here, the combination of CNN and GNN is a very promising solution, like in the [206]. First, the two-layer CNN was applied to the creation of the feature maps, and then two GNN layers were used to selectively filter out the discriminative features.

Recently, also Sinusoidal Representation Networks so-called SIRENs were applied to image segmentation. The essence of this approach is based on periodic activation functions for implicit neural representations. In fact, this AI solution mostly applied the sine periodic activation function. In [208] was proposed to analyze images, and in [209, 210] to segment medical images (cardiac MRI). However, this approach in the field of medical image segmentation requires still improvement.

Another interesting algorithm for natural image segmentation with was recently developed (April 2023) by Meta is the Segmentation Anything Model (SAM) [89, 90]. This AI-based algorithm enables cutting out any object from the image with a single click. It uses CNNs and transformer-based architectures for image processing, in particular, transformers-based architectures are applied to extract the features, compute the embedding, and pomp the encoder. The first attempt has been made to apply it in the field of medical imaging, however, in medical segmentation, it is still not so accurate in comparison to other application fields [91, 92]. The imperfections of the SAM algorithm in the field of medical image segmentation are mainly connected to insufficient numbers of training data. In [93], the authors proposed to apply the Med SAM Adapter to overcome the above limitations. The pre-training method like Masked Autoencoder (MAE), Contrastive Embedding-Mixup (e-Mix), and Shuffled Embedding Prediction (ShED) was applied. There is a lot of work in the area of medical image segmentation using machine learning, but relatively little addresses the issue related to the network learning process itself (along with data, a key element in achieving high accuracy of the process) [94].

The comparative comparison of the neural network architectures, learning algorithms, and datasets that are applied in the field of image segmentation in medicine is shown in **Table 1**. It turned out that still the most commonly used (taking into account accuracy of prediction) in these areas are ANN, and CNN constructed with the perceptrons and LIF neuron models and BP learning rules. Thus, the most commonly used learning algorithms in medical image segmentation are still on the low level of biological plausibility. On the other hand, in other image segmentation, in particular, biologically plausible learning algorithms are applied, for example, in the field of the images of handwritten digits [77]. Thus, **Table 1** presents works that contain information about the neuron model, architecture and type of neural network, input and output parameters to the network, and type of learning algorithm.

**Table 1.** The comparison of the AI-based algorithms applied in medical image scan segmentation

| Network Type | Neuron model | Average Accuracy [%] | Data sets - training/testing/validation sets [%] or training/testing sets [%] | Input parameters | Learning rule | Biological plausibility | Ref. |
|---|---|---|---|---|---|---|---|

| ANN | Perceproton | 99.10 | mammography images *lack of information* | mammography images – 33 features extracted by Region of Interest (ROI) | BP | low | [95] |
|---|---|---|---|---|---|---|---|
| CNN | Perceproton | 98.70 | Brain tumor, MRI color images *70/15/15* | MRI image scan, 12 features (mean, standard deviation (SD), entropy, Energy, contract, homogeneity, correlation, variance, covariance, root mean square (RMS), skewness, kurtosis) | BP | low | [96] |
| CNN | Perceproton | 96.00 | Echocardiograms *60/40* | Disease classification, cardiac chamber segmentation, viewpoints classification in echocardiograms | lack of information | low | [97] |
| CNN | Perceproton | 94.58 | brain tumor images *50/25/25* | brain tumor images | lack of information | low | [98] |
| CNN | Perceproton | 91.10 | simultaneous IVUS and OCT images | IVUS and OCT images | lack of information | low | [99] |
| CNN | Perceproton | 98.00 | 2-D ultrasound *49/49/2* | Classification of the cardiac view into 7 classes | lack of information | low | [100] |
| CNN | Perceproton | 93.30 | coronary cross-sectional images *80/20* | Detection of motion artifacts in coronary CCTA, classification of coronary cross-sectional images | lack of information | low | [101] |
| CNN | Perceproton | 99.00 | MRI image scan *60/40* | Bounding box localization of LV in short-axis MRI slices | lack of information | low | [102] |
| CNN and doc2vec | Perceproton | 96.00 | continuous wave Doppler cardiac valve images *94/4/2* | Automatic generation of text for continuous wave Doppler cardiac valve images | lack of information | low | [103] |
| Deep CNN + complex data preparation | Perceproton | 97.00 | Vessel segmentation *lack of information* | proposing a supervised segmentation technique that uses a deep neural network. Using structured prediction | lack of information | low | [104] |
| CNN and Transformer encoders | Perceproton | 90.70 | Automated Cardiac Diagnosis Challenge (ACDC), CT image scans from Synapse *60/40* | CT image scans | BP | low | [105] |
| CNN and Transformer encoders | Multi-layer Perceproton | 77.48 (Dice coefficient) | Multi-organ segmentation *lack of information* | CT image scans - Synapse multi-organ segmentation dataset | BP | low | [189] |
| CNN and Transformer encoders | Perception | 78.41 (Dice coefficient) | Multi-organ segmentation *lack of information* | CT image scans | BP | low | [181] |
| CNN, and RNN | Perceproton | 95.24 (REsNet50) 97.18(IncepnetV3) 98.03 (Dense-Net) | MRI image scan of the brain *80/20* | MRI image scan of the brain, modality, mask images | BP | low | [106] |

| | | | | | | | | |
|---|---|---|---|---|---|---|---|---|
| CNN, and RNN | Perceproton | 95.74 (REs-Net50) 97.14 (Dark-Net-53) | skin image *lack of information* | skin image | BP | low | [107] |
| SNN | LIF | 81.95 | baseline T1-weighted whole brain MRI image scan *lack of information* | The hippocampus section of the MRI image scan | ANN-SNN conversion | low | [108] |
| SNN | LIF | 92.89 | burn images *lack of information* | 256 × 256 burn image encoded into 24 × 256 × 256 feature maps | BP | low | [109] |
| SNN | LIF | 89.57 | skin images (melanoma and non-melanoma) *lack of information* | skin images converted into spikes using Poisson distribution | surrogated gradient descent | low | [110] |
| SNN | LIF | 99.60 | MRI scan of brain tumors *80/10/10* | 2D MRI scan of brain tumors | YO-LO-2-based transfer learning | low | [111] |
| SNN | LIF | 95.17 | microscopic images of breast tumor *lack of information* | microscopic images of breast tumor | Spike-Prop | low | [112] |
| GAN | Perceptron | 83.70 (Dice coefficient) DRIVE dataset 82.70 (Dice coefficient) STARE dataset | segmentation of retinal vessels *lack of information* | dataset for retinal vessel segmentation: DRIVE dataset and STARE dataset | BP | low | [191] |
| GAN | Perceptron | 94.60 | segmentation of the blood vessels of the retinal and the coronary and for the knee cartilage *lack of information* | dataset for retinal vessel segmentation: DRIVE dataset and coronary dataset | BP | low | [193] |
| GAN | Perceptron | 90.71 (Dice coefficient) | Brain segmentation data brain-MRI dataset *80/20* | Brain MRI image scan | BP | low | [192] |

*Legend:*
*BP – back propagation,*
*ANN – Artificial Neural Network,*
*SNN – Spiking Neural Network,*
*YOLO – you only look once algorithm,*
*spike-prop – supervised learning rule akin to traditional error-back-propagation for a network of spiking neurons with reasonable post-synaptic potentials,*
*MRI – Magnetic resonance imaging,*
*OCT– optical coherence tomography,*
*IVUS – intravascular ultrasound,*
*CCTA – Coronary Computed Tomography Angiography,*
*LV – left ventricle,*
*CT – computer tomography,*
*T1 weighted image – the basic pulse sequences in MRI, it shows the differences in the T1 relaxation times of tissue (T1 relaxation measures of how quickly the net magnetization vector recovers to its ground state),*
*GAN – Generative Adversarial Networks.*

The segmented structures (in this case organs and their disorders) may be next applied to the development of the 3D virtual environment [105]. These 3D objects may be implemented through for example, holograms displayed in the head-mounted display (HDMs) like Mixed Reality glasses in medical diagnostics [113], pre-operative imaging [114], surgical assistance [115, 116], robotics surgery [117], and medical education [81, 82]. However, the crucial issue is connected with the quality of obtained segmented structures, and this process can be significantly accelerated and improved by the use of Artificial Intelligence.

The crucial issue when the AI-based system is developed is connected with the accuracy and performance of algorithms. Many metrics have been introduced in image segmentation that enable the evaluation of algorithms. They can be split into two metric types, binary which takes into account two types of classification, and multi-class classification, in which the number of classes is higher than two. These metrics have been widely described in [163, 164]. The most commonly used metrics in medical image segmentation are the binary classifier $F-Score$ (F-measure) or so-called $F1-Score$ (also called Dice Coefficient) [165], Mean Absolute Error ($MAE$) [166], Mean Squared Error ($MSE$), Root-Mean-Squared Error ($RMSE$), Area Under Receiver Operator Curve ($AUROC$) [167] as well as Index of Union ($IoU$) [168].

## 7. Data availability

One of the key issues in the development of AI algorithms in the field of medicine is the availability and quality of data, i.e. access to electronic health records (EHRs) [118, 119]. Thus, the medical data should be anonymized. In **Table 2** a summary of publicly available retrospective image scan medical databases is presented. Some authors also provide anonymized data upon request. It is worth stressing that data, including medical image scans, are subjected to various types of biases [120]. The databases listed in Table 2 do not contain precise information regarding, for example, the ethnic composition of the study participants. Their age ranges and gender are usually disclosed. Moreover, another important issue concerning medical data is connected with Internet segmentation errors as was pointed out in the [169]. The authors discovered that the publicly available dataset contains duplicate records, which may contribute to the overlearning of some patterns in AI and ML models as well as result in false predictions. Also, the random procedure of splitting the database into raining and tasting sets will then influence the results obtained. As a consequence may lead to obtaining inflated classification.

**Table 2**. A summary of publicly available retrospective image scan medical databases.

| Database | Data source | Data type | Amount of data | Availability |
|---|---|---|---|---|
| Physionet | [121] | EEG, x-ray images, polysomnographic, | *Auditory evoked potential EEG-Biometric dataset* – 240 measurements from 20 subjects<br>*The Brno University of Technology Smartphone PPG Database (BUT PPG)* – 12 polysomnographic recordings<br>*CAP Sleep Database* - 108 polysomnographic recordings<br>*CheXmask Database: a large-scale dataset of anatomical segmentation masks for chest x-ray images* – 676 803 chest radiographs<br>*Electroencephalogram and eye-gaze datasets for robot-assisted surgery performance evaluation* – EEG from 25 subjects<br>*Siena Scalp EEG Database* – EEG from 14 subjects | Publics |
| Physionet | [121] | EEG, x-ray images, polysomnographic, | *Computed Tomography Images for Intracranial Hemorrhage Detection and Segmentation* – 82 CT After Traumatic Brain Injury (TBI)<br>*A multimodal dental dataset facilitating machine learning research and clinic service* -574 CBCT images from 389 patients<br>*KURIAS-ECG: a 12-lead electrocardiogram database with standardized diagnosis ontology*- EEG 147 subjects<br>*VinDr-PCXR: An open, large-scale pediatric chest X-ray dataset for interpretation of common thoracic diseases* – adult chest radiography (CXR) 9125 subjects<br>*VinDr-SpineXR: A large annotated medical image dataset for spinal lesions detection and classification from radiographs* - 10466 spine X-ray images from 5000 studies | Restricted access |
| National Sleep Research Resource | [122] | Polysomnography | *Apnea Positive Pressure Long-term Efficacy Study* – 1516 subject<br>*Efficacy Assessment of NOP Agonists in Non-Human Primates* – 5 subjects<br>*Maternal Sleep in Pregnancy and the Fetus* – 106 subjects<br>*Apnea, Bariatric surgery, and CPAP study* – 49 subjects<br>*Best Apnea Interventions in Research* – 169 subjects<br>*Childhood Adenotonsillectomy Trial* – 1243 subjects<br>*Cleveland Children's Sleep and Health Study* – 517 subjects<br>*Cleveland Family Study* – 735 subjects<br>*Cox & Fell (2020) Sleep Medicine Reviews* – 3 subjects<br>*Heart Biomarker Evaluation in Apnea Treatment* – 318 subjects<br>*Hispanic Community Health Study / Study of Latinos* – 16415 subjects<br>*Home Positive Airway Pressure* – 373 subjects<br>*Honolulu-Asia Aging Study of Sleep Apnea* – 718 subjects<br>*Learn* – 3 subjects<br>*Mignot Nature Communications* – 3000 subjects<br>*MrOS Sleep Study* – 2237 subjects<br>*NCH Sleep DataBank* – 3673 subjects<br>*Nulliparous Pregnancy Outcomes Study Monitoring Mothers-to-be* – 3012 subjects<br>*Sleep Heart Health Study* – 5804 subjects<br>*Stanford Technology Analytics and Genomics in Sleep* – 1881 subjects<br>*Study of Osteoporotic Fractures* – 461 subjects<br>*Wisconsin Sleep Cohort* – 1123 subjects | Publics on request (no commercial use) |
| Open Access Series of Imaging Studies - Oasis Brain | [123] | MRI Alzheimer's disease | OASIS-1 – 416 subjects<br>OASIS-2 – 150 subjects<br>OASIS-3 – 1379 subjects<br>OASIS-4 – 663 subjects | Publics on request (no commercial use) |
| openeuro | [124] | MRI, PET, MEG, EEG, and iEEG data (various types of disorders, depending on the database) | 595 MRI public datasets, 23 304 subjects<br>8 PET public datasets – 19 subjects<br>161 EEG public dataset – 6790 subjects<br>23 iEEG public dataset – 550 subjects<br>32 MEG public dataset – 590 subjects | Publics |
| brain tumor dataset | [125] | MRI, brain tumor | MRI - 233 subjects | Publics |
| Cancer Imaging Archive (TCIA) | [126] | MR, CT, Positron Emission Tomography, Computed Radiography, Digital Radiography, Nuclear Medicine, Other (a category | HNSCC-mIF-mIHC-comparison – 8 subjects<br>CT-Phantom4Radiomics – 1 subject<br>Breast-MRI-NACT-Pilot – 64 subjects<br>Adrenal-ACC-Ki67-Seg – 53 subjects<br>CT Lymph Nodes – 176 subjects<br>UCSF-PDGM – 495 subjects | Publics (Free access, registration required) |

| | | | used in DICOM for images that do not fit into the standard modality categories), Structured Reporting Pathology Various | UPENN-GBM – 630 subjects<br>Hungarian-Colorectal-Screening – 200 subjects<br>Duke-Breast-Cancer-MRI – 922 subjects<br>Pancreatic-CT-CBCT-SEG – 40 subjects<br>HCC-TACE-Seg – 105 subjects<br>Vestibular-Schwannoma-SEG – 242 subjects<br>ACRIN 6698/I-SPY2 Breast DWI – 385 subjects<br>I-SPY2 Trial – 719 subjects<br>HER2 tumor ROIs – 273 subjects<br>DLBCL-Morphology – 209 subjects<br>CDD-CESM – 326 subjects<br>COVID-19-NY-SBU – 1,384 subjects<br>Prostate-Diagnosis – 92 subjects<br>NSCLC-Radiogenomics – 211 subjects<br>CT Images in COVID-19 – 661 subjects<br>QIBA-CT-Liver-Phantom – 3 subjects<br>Lung-PET-CT-Dx – 363 subjects<br>QIN-PROSTATE-Repeatability – 15 subjects<br>NSCLC-Radiomics – 422 subjects<br>Prostate-MRI-US-Biopsy – 1151 subjects<br>CRC_FFPE-CODEX_CellNeighs – 35 subjects<br>TCGA-BRCA – 139 subjects<br>TCGA-LIHC – 97 subjects<br>TCGA-LUAD – 69 subjects<br>TCGA-OV – 143 subjects<br>TCGA-KIRC – 267 subjects<br>Lung-Fused-CT-Pathology – 6 subjects<br>AML-Cytomorphology_LMU – 200 subjects<br>Pelvic-Reference-Data – 58 subjects<br>CC-Radiomics-Phantom-3 – 95 subjects<br>MiMM_SBILab – 5 subjects<br>LCTSC – 60 subjects<br>QIN Breast DCE-MRI – 10 subjects<br>Osteosarcoma Tumor Assessment – 4 subjects<br>CBIS-DDSM – 1566 subjects<br>QIN LUNG CT – 47 subjects<br>CC-Radiomics-Phantom – 17 subjects<br>PROSTATEx – 346 subjects<br>Prostate Fused-MRI-Pathology – 28 subjects<br>SPIE-AAPM Lung CT Challenge – 70 subjects<br>ISPY1 (ACRIN 6657) – 222 subjects<br>Pancreas-CT – 82 subjects<br>4D-Lung – 20 subjects<br>Soft-tissue-Sarcoma – 51 subjects<br>LungCT-Diagnosis – 61 subjects<br>Lung Phantom – 1 subject<br>Prostate-3T – 64 subjects<br>LIDC-IDRI – 1010 subjects<br>RIDER Phantom PET-CT – 20 subjects<br>RIDER Lung CT – 32 subjects<br>BREAST-DIAGNOSIS – 88 subjects<br>CT COLONOGRAPHY (ACRIN 6664) – 825 sub-jects | |
|---|---|---|---|---|---|
| LUNA16 | [127] | CT, Lung Nodules | | LUNA16- 888 CT scans | Publics (Free access to all users) |
| MICCAI 2012 Prostate Challenge | [128] | MRI, Prostate Imaging | | Prostate Segmentation in Transversal T2-weighted MR images - Amount of Data: 50 training cases | Publics (Free access to all users) |
| IEEE Dataport | [129] | Ultrasound Images, Brain MRI, Ultra-widefield fluorescein angiography images, Chest X-rays, Mammograms, CT, Lung Image Database Consortium and Image, Thermal Images | | CNN-Based Image Reconstruction Method for Ultrafast Ultrasound Imaging: 31,000 images<br>OpenBHB: a Multi-Site Brain MRI Dataset for Age Prediction and Debiasing: >5,000 - Brain MRI.<br>Benign Breast Tumor Dataset: 83 patients - Mammograms.<br>X-ray Bone Shadow Suppression: 4,080 images<br>STROKE: CT series of patients with M1 thrombus before thrombectomy: 88 patients<br>Automatic lung segmentation results Nextmedproject - 718 of the 1012 LIDC-IDRI scans<br>PRIME-FP20: Ultra-Widefield Fundus Photography Vessel Segmentation Dataset -15 images<br>Plantar Thermogram Database for the Study of Diabetic Foot Complications - Amount of data: 122 subjects (DM group) and 45 subjects (control group) | A part Public and a part restricted (Subscription) |
| AIMI | [130] | Brain MRI studies, Chest X-rays, echocardiograms, CT | | BrainMetShare- 156 subjects<br>CheXlocalize: 700 subjects<br>BrainMetShare: 156 subjects | Publics (Free access) |

| Dataset | Ref | Modality | Details | Access |
|---|---|---|---|---|
| | | | COCA - Coronary Calcium and Chest CTs: Not specified<br>CT Pulmonary Angiography: Not specified<br>CheXlocalize: 700 subjects<br>CheXpert: 65,240 subjects<br>CheXphoto: 3,700 subjects<br>CheXplanation: Not specified<br>DDI - Diverse Dermatology Images: Not specified EchoNet-Dynamic: 10,030 subjects<br>EchoNet-LVH: 12,000 subjects<br>EchoNet-Pediatric: 7,643 subjects<br>LERA - Lower Extremity Radiographs: 182 subjects MRNet: 1,370 subjects<br>MURA: 14,863 studies Multimodal Pulmonary Embolism Dataset: 1,794 subjects<br>SKM-TEA: Not specified<br>Thyroid Ultrasound Cine-clip: 167 subjects<br>CheXpert:224,316 chest radiographs of 65,240 subjects | |
| fast MRI | [131] | MRI | fast MRI Knee: 1,500+ subjects<br><br>fast MRI Brain: 6,970 subjects<br><br>fast MRI Prostate: 312 subjects | Publics (Free access, registration required) |
| ADNI | [132] | MRI, PET | Scans Related to Alzheimer's Disease | Publics (Free access, registration required) |
| Pediatric Brain Imaging Dataset | [133] | MRI | Pediatric Brain Imaging Data-set Over 500 pediatric brain MRI scans | Publics (Free access to all users) |
| ChestX-ray8 | [134] | Chest X-ray Images | NIH Clinical Center Chest X-ray Dataset - Over 100,000 images from more than 30,000 subjects | Publics (Free access to all users) |
| Breast Cancer Digital Repository | [135] | MLO and CC images | BCDR-FM (Film Mammography-based Repository) - Amount of Data: 1010 subjects<br>BCDR-DM (Full Field Digital Mammography-based Repository)Amount of Data: 724 subjects | Publics (Free access, registration required |
| Brain-CODE | [136] | Neuroimaging | High-Resolution Magnetic Resonance Imaging of Mouse Model Related to Autism - 839 subjects | Restricted (Application for access is required and Open Data Releases) |
| RadImage-Net | [137] | PET, CT, Ultrasound, MRI with DICOM tags | 5 million images from over 1 million studies across 500,000 subjects | Publics subset available; Full dataset licensable; Academic access with restrictions |
| EyePACS | [138] | Retinal fundus images for diabetic retinopathy screening | Images for Training and validation set- 57,146 images Test set - 8,790 images | Available through the Kaggle competition |
| Medical Segmentation Decathlon | [139] | mp-MRI, MRI, CT | 10 data sets  Cases (Train/Test)<br>Brain        484/266<br>Heart         20/10<br>Hippocampus 263/131<br>Liver          131/70<br>Lung          64/32<br>Pancreas      282/139<br>Prostate       32/16<br>Colon         126/64<br>Hepatic Vessels 303/140<br>Spleen   41/20 | Open source license, available for research use |
| DDSM | [140] | Mammography images | 2,500 studies with images, subjects info - 2620 cases in 43 volumes categorized by case type | Publics (Free access) |
| LIDC-IDRI | [141] | CT Images with Annotations | 1018 cases with XML and DICOM files - Images (DICOM, 125GB), DICOM Metadata Digest (CSV, 314 kB), Radiologist Annotations/Segmentations (XML format, 8.62 MB), Nodule Counts by Patient (XLS), Patient Diagnoses (XLS) | Images and annotations are available for download with NBIA Data Retriever, usage under CC BY 3.0 |
| synapse | [142] | CT scans, Zip files for raw data, registration data | CT scans- 50 scans with variable volume sizes and resolutions<br>Labeled organ data -13 abdominal organs were manually labeled<br>Zip files for raw data - Raw Data: 30 training + 20 testing; Registration Data: 870 training-training + 600 training-testing pairs | Under IRB supervision, Available for participants |
| Mini-MIAS | [143] | Mammographic images | 322 digitized films on 2.3GB 8mm tape - Images derived from the UK National Breast Screening Programme and digitized with Joyce-Loebl scanning micro-densitometer to 50 microns, reduced to 200 microns and standardized to 1024x1024 pixels for the database | free for scientific research under a license agreement |
| Breast Cancer Histopathological Database (BreakHis) | [144] | microscopic images of breast tumor | 9,109 microscopic images of breast tumor tissue collected from 82 subjects | free for scientific research under a license agreement |
| Messidor | [145] | eye fundus color numerical images | 1200 eye fundus color numerical images of the posterior pole | free for scientific research under a license agreement |

## 8. Discussion and conclusions

The most commonly used neural networks in the field of medicine are CNNs and ANNs, see **Table 1**. Moreover, the combination of Transformers and CNNs as well as GAN allows users to achieve increasingly more accurate results, these methods require refinement. It is also worth noting that the diagnostics processes require the interpretation of visual scenes, and here GNNs-based solutions like scene graphs [200] and knowledge graphs [201] may be beneficial. It is also important to remember that GNNs are designed to perform tasks that neural networks like CNN cannot perform. SIRENs seem also to be an interesting solution. What was surprising was the fact that many works on the use of Machine Learning do not contain a detailed description of the neural network architecture and the description of learning, even the description of the data sets is very general (i.e. treats AI as a black box), which are key issues responsible for the accuracy and reliability of the approach used. The effectiveness of learning algorithms is compared among others in terms of the number of learning cycles, number of objective function calculations, number of floating-point multiplications, computation time, and sensitivity to local minima. In addition to the selection of appropriate parameters and network structure, the selection of an appropriate (effective) network learning algorithm is of key importance. The most commonly applied learning algorithm in ANNs is backpropagation, however, it has a rather slow convergence rate and as a consequence, ANN has more redundancy [146]. On the other hand, the training of the SNNs due to quite complicated dynamics and the non-differentiable nature of the spike activity remains a challenge [147]. The three types of ANN and SNN learning rules can be distinguished: unsupervised learning, indirect, supervised learning, and direct supervised learning. Thus, a commonly used learning algorithm in SNNs is the arithmetic rule SpikePropo, which is similar in concept to the backpropagation (BP) algorithm, in which network parameters are iteratively updated in a direction to minimize the difference between the final outputs of the network and target labels [148, 149]. The main difference between SNNs and ANNs is output dynamics. However, arithmetic-based learning rules are not a good choice for building biologically efficient networks. Other learning methods have been proposed for this purpose, including bio-inspired algorithms like spike-timing-dependent plasticity [150], spike-driven synaptic plasticity [151], and the tempotron learning rule [65, 76, 77]. STDP is unsupervised learning, which characterizes synaptic changes solely in terms of the temporal contiguity of presynaptic spikes and postsynaptic potentials or spikes [152], while spike-driven synaptic plasticity is supervised learning and uses rate coding. However, still, ANN with BP learning achieves a better classification performance than SNNs trained with STDP. To obtain better performance the combination of layer-wise STDP-based unsupervised and supervised spike-based BP was proposed [153, 154]. Other commonly used learning algorithms are ReSuMe [57], and Chronotron [58]. The tempotron learning rule implements gradient-descent dynamics, which minimizes a cost function that measures the amount by which the maximum voltage generated by erroneous patterns deviates from the firing threshold. Tempotron learning is efficient in learning spiking patterns where information is embedded in precise timing spikes (temporal coding). Instead, [155] proposed a neuron normalization technique and an explicitly iterative neuron model, which resulted in a significant increase in the SNNs' learning rate. However, training the network still requires a lot of labeled samples (input data). Another learning algorithm is indirect. It firstly trains ANN (created with perceptron's) and thereupon transforms it into its SNN version with the same network structure (i.e., ANN-SNN conversion) [156]. The disadvantage of such learning is the fact, that reliably estimating frequencies requires a nontrivial passage of time, and this learning rule fails to capture the temporal dynamics of a spiking system. The most popular direct supervised learning is gradient descent, which uses the first-spike time to encode input [157]. It uses the first-spike time to encode input signals and minimizes the difference between the network output and desired signals, the whole process of which is similar to the traditional BP. Thus, the application of the temporal coding-based learning rule, which could potentially carry the same information efficiently using less number of spikes than the rate coding, can help to increase the speed of calculations. On the other hand, active learning methods, including bio-inspired active learning (BAL), bio-inspired active learning on Firing Rate (BAL-FR), and bio-inspired active learning on membrane potential (BAL-M) have been proposed to reduce the size of the input data [158]. During the learning procedure, the labeled data sets are used to train the empirical behaviors of patterns, while the generalization behavior of patterns is extracted from unlabeled data sets. It leverages the difference between empirical and generalization behavior patterns to select the samples unmatched by the known patterns. This approach is based on the behavioral pattern differences of neurons in SNNs for active sample selection, and can effectively reduce the sample size required for SNNs training.

The impact of a bio-inspired AI-based system in clinical practice has significant potential for clinicians and medical experts. As can be clearly observed, further directions of development of Artificial Intelligence are leaning towards elaboration of not treating it as a black box and the development of Biological Artificial Intelligence, i. e. using neuron models that accurately reproduce experimentally measured values, understanding how information is transmitted, encoded and processed in the brain and mapping it in learning algorithms. The main issue is how replicated the architecture of the human brain and the mechanisms governing it. Biologically realistic large-scale brain models require a huge number of neurons as well as connections between them. Estimation of the behavior of a neuron network requires accurate models of the individual neurons along with accurate characterizations of the connections among them. In general, these models should contain all essential qualitative mechanisms and should provide results consistent with experimental physiological data. To fully characterize and predict the behavior of an identified network, one would need to know this architecture as well as any external currents or driving forces, and afferent input, applied to this network. Thus, the information transmission efficiency essentially depends on how neurons cooperate in the transfer process. The specific network architecture i.e. presence and distribution of long-range connections and the influence of inhibitor neurons, in particular the appropriate balance between excitatory and inhibitory neurons, make information transmission more effective [170]. Taking all these factors into account will put insight into the understanding of what factors contribute to the fact that the human brain is such a perfect computing machine. Then these mechanisms can be translated into the improvement of AI methods [171]. Moreover, it will put insight into the development of the next-generation AI, including Autonomous AI (AAI) as well as the development of brain simulators that balance computational complexity, energy efficiency, biological plausibility, and intellectual competence.

The second issue is connected with the so-called open data policy [211]. However, the publicly available datasets are not numerous, very often not labeled, described very generally, subject to bias, and additionally burdened with segmentation errors.

Another trend, which can be also observed is connected with the compliance of Artificial Intelligence with human rights, bioethics principles, and universal human values, which is especially important in medicine. For example, in Germany, a patient must give informed consent to the use of AI in the process of his diagnosis and treatment, which we believe is a good practice. Also, rules that should be fulfilled by the AI-based system like the Assessment List for Trustworthy Artificial Intelligence (ALTAI) [173-175], were formulated. In [163, 176] 10 ethical risk points (ERP) important to institutions, policymakers, teachers, students, and patients, including potential impacts on design, content, delivery, and AI-human-communication in the field of AI and Metaverse-based medical education were defined. Moreover, links are made between technical risks and ethical risks have been made. Now procedures need to be developed to enable their practical enforcement.

Thus, the integration of AI and Metaverse is a fact and suggests that AI may become the dominant approach for image scan segmentation and intelligent visual-content generation in the whole virtual world, not just medical applications [6, 159]. Recently, the Segment Anything Model (SAM) based on AI was introduced for natural images [89], in [160] SAM was proposed to be applied to medical images with a high level of accuracy. Better image segmentation contributes the higher-quality virtual objects. AI application in the context of the Metaverse is connected with the identification and categorization of meta-verse virtual items [161]. Moreover, AI may lead to more efficient cybersecurity solutions in the virtual world [162]. However, this is closely related to the accuracy of AI-based algorithms and, consequently, the accuracy of their training.

**Supplementary Materials:** Not applicable.

**Author Contributions:** "Conceptualization, A.P., and J.S.; methodology, A.P., J.S., and Z.R; software, A.P., Z. R.; formal analysis, A.P., J.S., and Z.R; investigation, A.P., J.S., and Z.R; resources, A.P. and Z.R.; data curation, A.P., and Z.R.; writing—original draft preparation, A.P., and Z.R.; writing—review and editing, A.P., J.S., and Z.R; visualization, A.P., and Z.R.; supervision, A.P.; project administration, A.P.; funding acquisition, J.S. All authors have read and agreed to the published version of the manuscript."

**Funding:** "This research received no external funding".

**Data Availability Statement:** Not applicable.

**Acknowledgments:** This study was partially supported by the National Centre for Research and Development (research grant Infostrateg I/0042/2021-00).

**Conflicts of Interest:** "The authors declare no conflict of interest."